\documentclass[aps,pra,superscriptaddress,twocolumn,showpacs]{revtex4}
\usepackage{amsfonts}
\usepackage{amsmath}
\usepackage{amssymb}
\usepackage{graphicx}

\begin{document}

\title{Structure and dynamics of a rotating superfluid Bose-Fermi mixture}
\author{Linghua Wen}
\email{linghuawen@ysu.edu.cn}
\affiliation{College of Science, Yanshan University, Qinhuangdao 066004, China}
\author{Jinghong Li}
\affiliation{College of Environment and Chemical Engineering, and College of Science,
Yanshan University, Qinhuangdao 066004, China}
\date{\today }

\begin{abstract}
We investigate the structure and dynamics of a rotating superfluid
Bose-Fermi mixture (SBFM) made of superfluid bosons and two-component (spin
up and down) superfluid fermions. A ground-state phase diagram for the
nonrotating case of a SBFM with specific parameters is given, where the
ground-state configuration of a nonrotating SBFM is mainly determined by the
boson-fermion interaction. For the rotating case of a SBFM with a
sufficiently large rotation frequency, we show that the system supports a
mixed phase and three typical layer separated phases. In particular, the
visible vortex formation in the fermionic superfluid exhibits a remarkable
hysteresis effect during the dynamical evolution of a rotating SBFM, which
is evidently different from the case of rotating two-component Bose-Einstein
condensates.
\end{abstract}

\pacs{67.85.Pq, 67.85.Lm, 03.75.Kk, 03.75.Lm}
\maketitle

\section{Introduction}

Superfluidity plays a key role in many fields of physics, such as liquid
helium \cite{Leggett}, ultracold atomic gases \cite{Pitaevskii,Pethick},
quantum magnets \cite{Zapf}, and astrophysics \cite{Liberati}. In
particular, superfluid Bose-Fermi mixtures (SBFMs) in ultracold atomic gases
have attracted considerable interest recently. It is well known that for
liquid helium the superfluidity of bosonic $^{4}$He and fermionic $^{3}$He
can be achieved separately. However, the simultaneous superfluidity for the
two isotopes is remarkably prevented by the strong interactions between the
two species in spite of such a superfluid $^{4}$He-$^{3}$He mixture being a
long-sought object \cite{Rysti}. By contrast, a superfluid Bose-Fermi
mixture (SBFM) in ultracold atomic gases can be realized by using the
combination of the Feshbach resonance and radio-frequency techniques. Most
recently, a SBFM consisting of condensed $^{7}$Li bosons and ultracold $^{6}$%
Li fermions in two-spin states has been produced by Salomon's group \cite%
{Ferrier-Barbut}. The experimental breakthrough provides new opportunities
to study the intriguing properties of SBFMs inaccessible in a Bose-Einstein
condensate (BEC) with arbitrary spin \cite{Ho1,YWu}, pure bosonic superfluid
mixtures, and pure fermionic superfluid mixtures. As a matter of fact, many
novel physical characteristics have been predicted theoretically in SBFMs,
including the saturation effect of nonlinear interaction in the
strong-coupling unitarity limit \cite{Adhikari1}, localized Bose-Fermi
bright soliton \cite{Adhikari2}, possibility of simulating dense quantum
chromodynamics matter \cite{Maeda}, phase transition overlapping with phase
separation \cite{Ramachandhran}, and the Faraday pattern generation \cite%
{Abdullaev}.

All existing studies of the SBFMs refer to the nonrotating case. Considering
that one of the most striking hallmarks of a superfluid is its response to
rotation, in this paper we investigate the combined effects of rotation and
nonlinear interatomic interaction on the exact two-dimensional (2D)
topological structure and dynamics of a rotating SBFM, focusing on the
superfluid $^{7}$Li-$^{6}$Li mixture case as the prototype. Here we present
a phenomenological dissipation model \cite{Wen1} combining with the unitary
Schr\"{o}dinger equation for an interacting SBFM \cite{Adhikari1} to
describe the dynamics of a rotating SBFM. A ground-state phase diagram for a
nonrotating SBFM with specific parameters is given. We find that the
ground-state structure of a nonrotating SBFM is mainly determined by the
ratio of the boson-fermion (BF) scattering length to the boson-boson (BB)
scattering length. Furthermore, it is shown that the rotating SBFM with a
sufficiently large rotation frequency can display four steady structures: a
mixed phase and three typical layer separated phases. In particular, we show
that the generation of visible vortices in the fermionic superfluid displays
an evident hysteresis effect during the time evolution of a rotating SBFM.
Due to the Cooper pair entity of fermionic superfluid and the presence of BF
and fermion-fermion (FF) interactions, it is demonstrated that the
topological structure and the dynamics of a rotating SBFM are evidently
different from the usual cases of rotating two-component Bose-Einstein
condensates (BECs).

The paper is organized as follows. In Sec. II, we describe a
phenomenological model for a rotating SBFM. In Sec. III, we study the
equilibrium structure of the rotating SBFM. A phase diagram for a
nonrotating SBFM is given, and the phase structures of the rotating SBFMs
with various parameter values are analyzed. In Sec. IV, we discuss the
dynamics of vortex formation in a rotating SBFM. The conclusion is outlined
in the last section.

\section{Model}

We consider a superfluid Bose-Fermi mixture composed of superfluid bosons
and two-component superfluid fermions with equal populations of spin up and
down, corresponding to the most favorable condition for Cooper pairing. The
superfluid Bose-Fermi gas is confined in a harmonic trap rotating around the
$z$ axis with angular velocity $\Omega $. In the present work we use a
combined phenomenological model based on a phenomenological dissipation
model \cite{Wen1} and a Galilei-invariant nonlinear unitarity model \cite%
{Adhikari1} to investigate the dynamics of the vortex formation and the
structure of the equilibrium state of a rotating SBFM. In this
phenomenological model, the order parameter $\Psi _{b}$ of a bosonic
superfluid (i.e., a Bose-Einstein condensate) and the order parameter $\Psi
_{p}$ of a fermionic superfluid \cite{Adhikari1} obey the coupled equations
in the rotating frame,%
\begin{eqnarray}
(i-\gamma )\hbar \frac{\partial \Psi _{b}}{\partial t} &=&\left[ -\frac{%
\hbar ^{2}\nabla ^{2}}{2m_{b}}+U_{b}+\mu _{b}(n_{b},a_{b})-\Omega L_{z}%
\right] \Psi _{b}  \notag \\
&&+G_{bp}\left\vert \Psi _{p}\right\vert ^{2}\Psi _{b},  \label{SBE} \\
(i-\gamma )\hbar \frac{\partial \Psi _{p}}{\partial t} &=&\left[ -\frac{%
\hbar ^{2}\nabla ^{2}}{2m_{p}}+U_{p}+\mu _{p}(n_{p},a_{f})-\Omega L_{z}%
\right] \Psi _{p}  \notag \\
&&+G_{bp}\left\vert \Psi _{b}\right\vert ^{2}\Psi _{p},  \label{SFE}
\end{eqnarray}%
where for simplicity we assume that the degree of dissipation $\gamma $ of
the fermionic superfluid is the same as that of the BEC. $m_{b}$ is the mass
of a bosonic atom, $m_{p}=2m_{f}$ is the mass of a Cooper pair with $m_{f}$
being the mass of a single fermion, and $L_{z}=i\hbar (y\partial
_{x}-x\partial _{y})$ denotes the $z$ component of the angular-momentum
operator. $U_{b}=m_{b}(\omega _{b}^{2}r^{2}+\omega _{zb}^{2}z^{2})/2$ is the
external trapping potential for the bosons, $U_{p}=m_{p}(\omega
_{f}^{2}r^{2}+\omega _{zf}^{2}z^{2})/2$ is the trapping potential for the
fermionic superfluid, and $\omega _{b}$($\omega _{f}$) and $\omega _{zb}$($%
\omega _{zf}$) are the radial trap frequency and the axial trap frequency
for the bosons (fermions), respectively. $G_{bp}=4\pi \hbar
^{2}a_{bf}/m_{bf} $, where $a_{bf}$ is the BF $s$-wave scattering length,
and $m_{bf}=m_{b}m_{f}/(m_{b}+m_{f})$ is the BF reduced mass. The bulk
chemical potential of the BEC is given by%
\begin{equation}
\mu _{b}(n_{b},a_{b})=\frac{\hbar ^{2}}{m_{b}}n_{b}^{2/3}f(n_{b}^{1/3}a_{b}),
\label{ChemPotentialBoson}
\end{equation}%
where
\begin{equation}
f(x)=\frac{4\pi \left( x+\alpha x^{5/2}\right) }{1+\lambda x^{3/2}+\beta
x^{5/2}},  \label{BosonFunction}
\end{equation}%
$n_{b}$ is the local density of bosons, and $a_{b}$ is the BB $s$-wave
scattering length. Here we choose $(\alpha -\lambda )=32/(3\sqrt{\pi })$, $%
\beta =4\pi \alpha /\eta $ with $\eta $ $=22.22$, $\alpha =32\xi /(3\sqrt{%
\pi })$, and $\lambda =32(\xi -1)/(3\sqrt{\pi })$ with $\xi =1.1$ \cite%
{Adhikari1}. The bulk chemical potential of the Fermi superfluid is
expressed as \cite{Adhikari1,Recati,Adhikari3}
\begin{equation}
\mu _{p}(n_{p},a_{f})=\frac{2\hbar ^{2}}{m_{p}}(6\pi
^{2}n_{p})^{2/3}g(2^{1/3}n_{p}^{1/3}a_{f}),  \label{ChemPotentialFermion}
\end{equation}%
where
\begin{equation}
g(x)=1+\frac{\delta x}{1-kx},  \label{FermionFunction}
\end{equation}%
$n_{p}=n_{f}/2$ is the local density of Cooper pairs with $n_{f}$ being the
total local density of fermions, $a_{f}$ denotes the attractive FF
scattering length, $\delta =20\pi /(3\pi ^{2})^{2/3}$, and $k=\delta
/(1-\zeta )$ with $\zeta =0.44$. The choices of the parameters in Eqs. (\ref%
{BosonFunction}) and (\ref{FermionFunction}) are consistent with the
unitarity and the Lee-Yang-Huang limits \cite{Adhikari1,Lee1,Lee2} as well
as the relevant results of the Monte Carlo calculations \cite%
{Blume1,Carlson,Astrakharchik,Blume2,Chang}.

This phenomenological model is a variation of that in \cite%
{Tsubota,Kasamatsu1} and a generalization of that of a rotating
single-component BEC \cite{Wen1}, and it has good predictive power \cite%
{Wen2}. The largest merit of the phenomenological dissipation model is that
by virtue of this model one can obtain not only\ the steady states of a
rotating system but also reveal the dynamics of vortex formation. In
addition, the present model is valid from weak coupling to unitarity for
both bosons and fermions \cite{Adhikari1}.

As a matter of fact, the unitary Schr\"{o}dinger model for nonrotating
superfluid fermions was presented in Refs. \cite%
{Adhikari1,Adhikari3,Salasnich}, where the bulk chemical potential was given
by Eqs. (\ref{ChemPotentialFermion}) and (\ref{FermionFunction}) (see also
Eqs. (10), (11) and (28) in Ref. \cite{Adhikari1}). The lowest-order term $%
g(x)=1$ in Eq. (\ref{FermionFunction}) results in the bulk chemical
potential of a Fermi superfluid in the absence of FF interaction ($a_{f}=0$%
). Furthermore, the next-order term $g(x)=1+\delta x$ can lead to a known
analytical result in the small-gas-parameter regime (medium value FF
scattering length $a_{f}$) as obtained in Ref. \cite{Lee1}. When $%
n_{p}^{1/3}a_{f}$ tends to negative infinity (the so-called unitary limit),
the function $g(x)$ obviously approaches an asymptotic value $(1-\delta /k)$%
. Thus the function (\ref{ChemPotentialFermion}) with Eq. (\ref%
{FermionFunction}) provides a smooth interpolation between the bulk chemical
potential of a Fermi superfluid in the weak-coupling limit and that in the
unitary limit for both the uniform case and the trapped case \cite%
{Adhikari1,Adhikari3,Lee1}. It was shown that the results with the choice of
the fitting parameters $\delta =20\pi /(3\pi ^{2})^{2/3}$, $k=\delta
/(1-\zeta ),$ and $\zeta =0.44$ \cite{Adhikari1} agreed well with the
corresponding Monte Carlo data \cite{Blume2,Chang}. Therefore, we will use
the choice in the present work.

Next, we consider the two-dimensional problem by assuming the translation
invariance along the $z$ axis (i.e., $\omega _{zb}=\omega _{zf}=0$),
reducing the order parameters as $\Psi _{i}(\overrightarrow{r},t)=\psi
_{i}(x,y,t)/\sqrt{R_{z}}(i=b,p)$ with the typical size $R_{z}$ along the $z$
axis. After a straight-forward calculation, we obtain 2D coupled equations,%
\begin{eqnarray}
(i-\gamma )\hbar \frac{\partial \psi _{b}}{\partial t} &=&\left[ -\frac{%
\hbar ^{2}}{2m_{b}}\left( \nabla _{x}^{2}+\nabla _{y}^{2}\right) +F(\psi
_{b})-\Omega L_{z}\right] \psi _{b}  \notag \\
&&+\left[ \frac{m_{b}\omega _{b}^{2}}{2}\left( x^{2}+y^{2}\right) +\frac{%
G_{bp}}{R_{z}}\left\vert \psi _{p}\right\vert ^{2}\right] \psi _{b},
\label{2DSBE} \\
(i-\gamma )\hbar \frac{\partial \psi _{p}}{\partial t} &=&\left[ -\frac{%
\hbar ^{2}}{2m_{p}}(\nabla _{x}^{2}+\nabla _{y}^{2})+J(\psi _{b})-\Omega
L_{z}\right] \psi _{p}  \notag \\
&&+\left[ m_{f}\omega _{f}^{2}(x^{2}+y^{2})+\frac{G_{bp}}{R_{z}}\left\vert
\psi _{b}\right\vert ^{2}\right] \psi _{p},  \label{2DSFE}
\end{eqnarray}%
where%
\begin{equation}
F(\psi _{b})=\frac{4\pi \hbar ^{2}\left( a_{b}R_{z}^{-1}\left\vert \psi
_{b}\right\vert ^{2}+\alpha a_{b}^{5/2}R_{z}^{-3/2}\left\vert \psi
_{b}\right\vert ^{3}\right) }{m_{b}\left( 1+\lambda
a_{b}^{3/2}R_{z}^{-1/2}\left\vert \psi _{b}\right\vert +\beta
a_{b}^{5/2}R_{z}^{-5/6}\left\vert \psi _{b}\right\vert ^{5/3}\right) },
\label{Ffunction}
\end{equation}%
and
\begin{equation}
J(\psi _{p})=\frac{2\hbar ^{2}}{m_{p}}(6\pi ^{2})^{2/3}\left\vert \psi
_{p}\right\vert ^{4/3}\left( 1+\frac{2^{1/3}\delta a_{f}\left\vert \psi
_{p}\right\vert ^{2/3}}{1-2^{1/3}ka_{f}\left\vert \psi _{p}\right\vert ^{2/3}%
}\right) .  \label{Jfunction}
\end{equation}%
The initial order parameters are normalized as $N_{i}=\iint \psi
_{i}(x,y,t=0)dxdy$ $(i=b,p)$ with $N_{b}$ being the initial number of bosons
and $N_{p}$ being the initial number of Cooper pairs ($N_{p}=N_{f}/2$ and $%
N_{f}$ is the total number of fermions). By introducing the notations $%
\omega _{bf}=(\omega _{b}+\omega _{f})/2$, $d_{0}=\sqrt{\hbar
/(2m_{bf}\omega _{bf})}$, $x_{0}=x/d_{0}$, $y_{0}=y/d_{0}$, $t_{0}=t\omega
_{bf}$ and $\Omega _{0}=\Omega /\omega _{bf}$, and replacing the wave
functions as $\psi _{i}\rightarrow \sqrt{N_{i}}\psi _{i}/d_{0}$, we obtained
the rescaled dimensionless 2D equations for the mixture%
\begin{eqnarray}
(i-\gamma )\frac{\partial \psi _{b}}{\partial t} &=&\left[ -\frac{m_{bf}}{%
m_{b}}(\nabla _{x}^{2}+\nabla _{y}^{2})+C_{1}\left\vert \psi _{p}\right\vert
^{2}-\Omega L_{z}\right] \psi _{b}  \notag \\
&&+\frac{4\pi \hbar }{m_{b}\omega _{bf}}\frac{\left( A_{1}\left\vert \psi
_{b}\right\vert ^{2}+\alpha B_{1}\left\vert \psi _{b}\right\vert
^{4/3}\right) }{1+\lambda D_{1}\left\vert \psi _{b}\right\vert +\beta
E_{1}\left\vert \psi _{b}\right\vert ^{5/3}}\psi _{b}  \notag \\
&&+\frac{m_{b}\omega _{b}^{2}}{4m_{bf}\omega _{bf}^{2}}\left(
x^{2}+y^{2}\right) \psi _{b},  \label{DimLess2DSBE} \\
(i-\gamma )\frac{\partial \psi _{p}}{\partial t} &=&\left[ -\frac{m_{bf}}{%
m_{p}}(\nabla _{x}^{2}+\nabla _{y}^{2})+C_{2}\left\vert \psi _{b}\right\vert
^{2}-\Omega L_{z}\right] \psi _{p}  \notag \\
&&+A_{2}\left\vert \psi _{p}\right\vert ^{4/3}\left( 1+\frac{\delta
B_{2}\left\vert \psi _{p}\right\vert ^{2/3}}{1-kB_{2}\left\vert \psi
_{p}\right\vert ^{2/3}}\right) \psi _{p}  \notag \\
&&+\frac{m_{f}\omega _{f}^{2}}{2m_{bf}\omega _{bf}^{2}}\left(
x^{2}+y^{2}\right) \psi _{p},  \label{DimLess2DSFE}
\end{eqnarray}%
where the number subscript $0$ is omitted for simplicity. The corresponding
coefficients are $A_{1}=a_{b}N_{b}d_{0}^{-2}R_{z}^{-1}$, $%
B_{1}=a_{b}^{5/2}N_{b}^{3/2}d_{0}^{-3}R_{z}^{-3/2}$, $C_{1}=8\pi
a_{bf}N_{p}R_{z}^{-1}$, $D_{1}=a_{b}^{3/2}N_{b}^{1/2}d_{0}^{-1}R_{z}^{-1/2}$%
, $E_{1}=a_{b}^{5/2}N_{b}^{5/6}d_{0}^{-5/3}R_{z}^{-5/6}$, $A_{2}=(2\hbar
/m_{p}\omega _{bf})(6\pi ^{2}N_{p}d_{0}^{-2})^{2/3}$, $%
B_{2}=(2N_{p}d_{0}^{-2})^{1/3}a_{f}$, and $C_{2}=8\pi a_{bf}N_{b}R_{z}^{-1}$.

In the following, we numerically solve the 2D coupled equations (\ref%
{DimLess2DSBE}) and (\ref{DimLess2DSFE}) which requires an enormous
computational effort. The initial ground-state order parameters $\psi
_{b}(x,y,t=0)$ and $\psi _{p}(x,y,t=0)$ of the system can be obtained by the
imaginary-time propagation method \cite{Zhang,Xu,Wen3} based on the
Peaceman-Rachford method \cite{Peaceman,Wen4}. Here we consider a superfluid
$^{7}$Li-$^{6}$Li mixture in harmonic traps satisfying $U_{b}(x,y)=m_{b}%
\omega _{b}^{2}\left( x^{2}+y^{2}\right) =U_{f}(x,y)=m_{f}\omega
_{f}^{2}\left( x^{2}+y^{2}\right) $, with $m_{b\text{ }}$being the mass of $%
^{7}$Li atom and $m_{f\text{ }}$ being the mass of the $^{6}$Li atom. The
parameters are chosen as $\omega _{b}=2\pi \times 100$ Hz, $R_{z}=10$ $\mu $%
m, and $\gamma =0.03$, corresponding to a temperature of about $0.1T_{c}$
\cite{Choi}. Recently, a mixture of $^{7}$Li superfluid and $^{6}$Li
superfluid has been realized by Salomon's group \cite{Ferrier-Barbut}. Thus
the above assumption is valid and feasible, and the relevant results can be
tested under the current experimental conditions.

\section{Steady structure of a rotating SBFM}

For convenience, we introduce two relative interaction strengths, $%
R_{1}=a_{bf}/a_{b}$ and $R_{2}=a_{f}/a_{b}$. The FF attractive scattering
length $a_{f}$ can be varied from zero to negative infinity, and the
absolute value of the BF scattering length $\left\vert a_{bf}\right\vert $
is assumed to be not too large (otherwise, it requires special attention
\cite{Adhikari1}).

\begin{figure}[tbp]
\centerline{\includegraphics*[width=7cm]{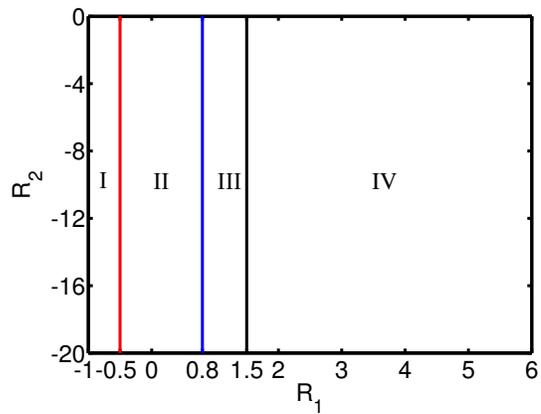}}
\caption{(Color online) Ground-state phase diagram of a nonrotating
superfluid $^{7}$Li-$^{6}$Li (Bose-Fermi) mixture, where $R_{1}=a_{bf}/a_{b}$
and $R_{2}=a_{f}/a_{b}$. The parameters are $N_{b}=1000$, $N_{p}=100$, and $%
a_{b}=50$ nm. Regions I--IV represent the system collapse regime, mixed
phase, layer-separated phase, and inlaid separated phase, respectively. }
\label{Figure1}
\end{figure}

Figure 1 shows the ground-state phase diagram of a static superfluid $^{7}$%
Li--$^{6}$Li mixture with fixed BB repulsive interaction. The corresponding
parameters are $a_{b}=50$ nm, $N_{b}=1000$, and $N_{p}=100$. There exist
four possible phases depending on the values of $R_{1}$ and $R_{2}$, and the
typical density profiles corresponding to phases II-IV are displayed in Fig.
2. In Fig. 1, region I denotes the system collapse regime. For a
sufficiently strong attractive BF interaction (here, $R_{1}<-0.5$, i.e., $%
a_{bf}<-25$ nm), the system undergoes a simultaneous collapse of the density
profile of the BEC and that of the fermionic superfluid. Physically, the
critical value of BF scattering length is governed by the balance between
the kinetic energy of bosons and Cooper pairs and the mutual attractive BF
interaction. When the BF attraction becomes sufficiently strong, it can no
longer be stabilized by the kinetic energy. Therefore the mixture lowers its
energy via increasing the densities of bosons and Cooper pairs, and finally
the bosonic superfluid or the fermionic one or both the bosonic and
fermionic superfluids collapse simultaneously due to instability. Region II
represents a miscible phase in which the density profile of the BEC has a
much larger spatial extension than that of the fermionic superfluid due to
the BB repulsion and the FF attraction [see Figs. 2(a1) and 2(a2)]. Region
III is a layer-separated phase, where the BEC is completely expelled outside
the fermionic superfluid because of the strong BF repulsive interaction
[Figs. 2(b1) and 2(b2)], which is usually referred as demixing. The similar
mixing-demixing transition phenomenon has also been found in the studies of
degenerate boson-fermion mixtures \cite{Molmer,Roth,Capuzzi,Adhikari4,Wen5}
or two-component BECs \cite{Wen4,Ho2,Pu,Trippenbach,LWen}. Finally, region
IV marks an inlaid separated phase in which the fermionic superfluid is
divided into two segments and inlaid into the outskirts of the BEC [Figs.
2(c1) and 2(c2)].

\begin{figure}[tbp]
\centerline{\includegraphics*[width=7.8cm]{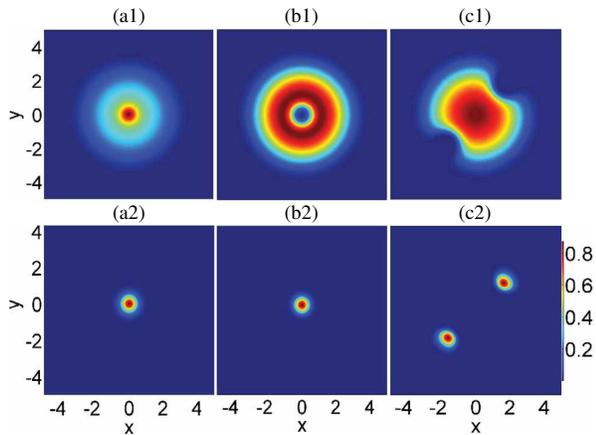}}
\caption{(Color online) Density profiles (a)--(c) correspond to phases
II--IV in Fig. 1, respectively. Here, $1$ labels the bosonic superfluid and $%
2$ denotes the fermionic one of the system. (a) $R_{1}=-0.4,R_{2}=-20,$ (b) $%
R_{1}=1,R_{2}=-20,$ and (c) $R_{1}=6,R_{2}=-20$. The other parameters are
the same as those in Fig. 1. The darker color area indicates the lower
density. $x$ and $y$ are in units of $d_{0}$. }
\label{Figure2}
\end{figure}

The ground-state structures of SBFMs are different from those of degenerate
boson-fermion mixtures \cite{Molmer,Roth,Capuzzi,Adhikari4,Wen5}. In the
latter case, there is no $s$-wave interaction between identical fermions in
the spin-polarized state due to the Pauli exclusion principle. Depending on
the BB and BF interactions, the density profile of a degenerate
boson-fermion mixture may display a core-shell-shaped separated phase, where
the Fermi gas forms a shell around or a core inside the BEC, or even both
\cite{Molmer,Roth,Capuzzi,Adhikari4,Wen5}. However, our simulation shows
that in a wide range of parameter values, the SBFM supports neither a
separated phase of fermions constituting a shell around the BEC nor a
staggered separated phase of fermions becoming both a shell around and a
core inside the BEC (see Figs. 1 and 2); even the total number of fermions
is much larger than that of bosons. In addition, for a fixed value of $R_{1}$
(i.e., a fixed BF interaction), the variation of the parameter $R_{2}$ (FF
interaction) dose not influence the phase structure of the SBFM (Fig. 1).
Furthermore, when the repulsive BF interaction of a SBFM is sufficiently
strong, the system exhibits an inlaid separated phase, where the fermionic
superfluid embeds into the periphery of the bosons by means of two fragment
superfluids [Figs. 2(c1) and 2(c2)]. It is known that the fundamental entity
of a fermionic superfluid is the Cooper pair \cite%
{Ramachandhran,Adhikari5,Gubbels} and there exists attractive FF interaction
between the fermion with spin up and that with spin down. Thus the
center-of-mass momentum of each Cooper pair in the fermionic superfluid
remains zero, irrespective of the concrete value of $a_{f}$, which may
account for the above differences.

On the other hand, the equilibrium properties of the SBFMs are also
evidently different from those of two-component BECs \cite%
{Wen4,Ho2,Pu,Trippenbach,LWen}. For the case of two-component BECs, by
varying the particle numbers of the two BECs or the ratios of intra- and
intercomponent interaction strengths, one can obtain many symmetric
separated phases \cite{Ho2,Pu,LWen}\ and even asymmetric separated phases
\cite{Wen4,Trippenbach}. Physically, there exist three kinds of actual $s$%
-wave interactions (including BB interaction, BF interaction, and FF
interaction) in a SBFM. In particular, when the bosonic and fermionic
scattering lengths $a_{b}$ and $\left\vert a_{f}\right\vert $ tend to
infinity (i.e., in the strong-coupling unitarity limit), both the bosonic
and fermionic interactions exhibit unitarity saturation effects due to the
constraints of quantum mechanics \cite{Adhikari1}. In this context, the
quasiparticle of a Cooper pair in the fermionic superfluid can just be
considered as a \textquotedblleft composite boson\textquotedblright\ rather
than a single bosonic atom, which largely results in the differences of the
ground-state structures.

In Fig. 3, we display the steady density profiles $\left\vert \psi
_{b}\right\vert ^{2}$ (row $1$) and $\left\vert \psi _{p}\right\vert ^{2}$
(row $2$), and the corresponding phase profiles of $\psi _{b}$ (row $3$) and
$\psi _{p}$ (row $4$) at $t=500$ for a harmonic trap rotating with $\Omega
=0.94$. Here the value of the phase varies continuously from $0$ to $2\pi $,
and the end point of the boundary between a $2\pi $ phase line and a $0$
phase line denotes a phase defect (i.e., a vortex with anticlockwise
rotation). The density profiles of three initial states corresponding to the
three columns\ (left, middle, and right) in Fig. 3 are given in Figs.
2(a)--2(c), respectively.

\begin{figure}[tbp]
\centerline{\includegraphics*[width=8cm]{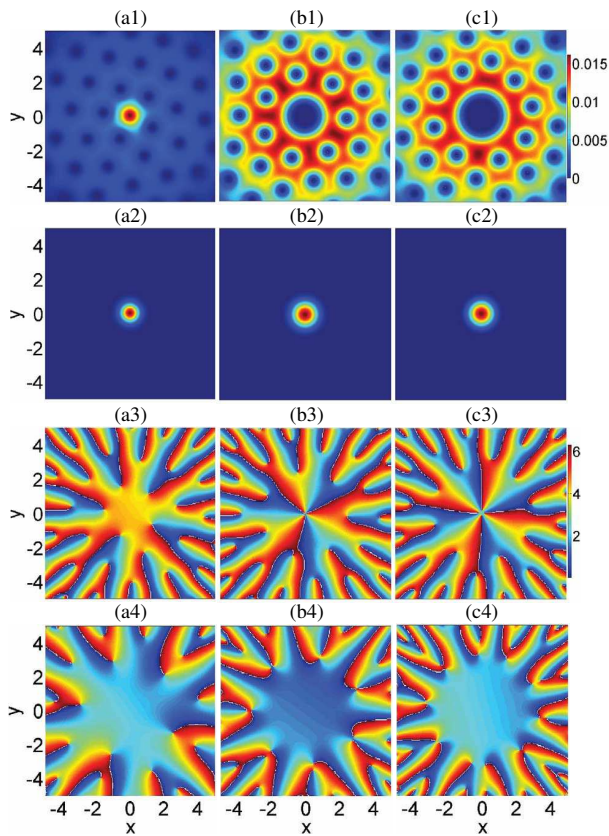}}
\caption{(Color online) Steady density profiles (the top two rows) and phase
profiles (the bottom two rows) at $t=500$ after rotating the superfluid
Bose-Fermi mixture with $\Omega =0.94$, \ where $1$ and $3$ denote the
bosonic superfluid, while $2$ and $4$ represent the fermionic superfluid.
The initial states corresponding to three cases of $a,b,$ and $c$ (from left
to right) are illustrated in Figs. 2(a1,a2), Figs. 2(b1,b2), and Figs.
2(c1,c2), respectively. The other parameters are the same as those in Figs.
1 and 2. The value of phase varies continuously from $0$ to $2\protect\pi $.
The darker color area indicates lower density or phase. Here, $x$ and $y$
are in units of $d_{0}$. }
\label{Figure3}
\end{figure}

As shown in Figs. 3(a1) and 3(a3), when the rotating SBFM with an initial
state of $R_{1}=-0.4$ and $R_{2}=-20$ reaches an equilibrium state, one can
see that there are several visible vortices \cite{Wen1,Wen2} constituting a
triangular lattice in the periphery of the bosonic superfluid. In the
meantime, the trap center is occupied by a density peak of bosons due to the
attractive BF interaction. However, no visible vortex can be generated in
the density of the fermionic superfluid even for large rotation frequency of
$\Omega =0.94$ [see Fig. 3(a2)]. From Fig. 3(a4), there exist some phase
defects that are located on the outskirts of the superfluid fermions. Since
these phase defects are invisible in the \textit{in situ} density profile of
fermionic superfluid and contribute to neither the angular momentum nor the
energy of the SBFM, there are referred to as ghost vortices \cite%
{Wen1,Tsubota,Kasamatsu1,Wen2,Wen4}. For the case of $R_{1}=1$ and $%
R_{2}=-20 $, there is a large circular density hole of bosons in the trap
center which looks like a giant vortex (a multiquantized vortex) and is
surrounded by an expected visible vortex lattice [Fig. 3(b1)]. In the phase
profile displayed in Fig. 3(b3), we find that there are three phase defects
in the trap center which are close to each other. The three singly quantized
phase defects show that the circular density hole is not a giant vortex.
Actually, they are known as singly quantized hidden vortices \cite%
{Wen1,Wen2,Wen4,Mithun} because they carry significant angular momentum,
though they are invisible in the \textit{in situ} density profile of the
BEC. Only after including the hidden vortices can the Feynman rule be
satisfied \cite{Wen1,Fetter,Cooper}. As seen in Figs. 3(b2) and 3(b4), the
fermionic superfluid with ghost vortices distributing on the outskirts of
the cloud is completely pulled inside the trap center (i.e., the region of
the large density hole of the BEC) due to the competition between the BF
repulsion and the rotation repulsion, which indicates a fully separated
phase of the rotating SBFM.

With the further increase of the BF repulsive interaction, a nonrotating
SBFM will develop into an inlaid separated configuration, where the
superfluid fermions lie on the outer edges of the bosons via the form of two
fragments [Figs. 2(c1) and 2(c2)]. Counterintuitively, we find that the
steady structure of a rotating SBFM with $R_{1}=6$ and $R_{2}=-20$ is
similar to that with $R_{1}=1$ and $R_{2}=-20,$ as shown in Figs.
3(c1)--3(c4). Here the fermionic superfluid is separately repelled to the
trap center region rather than the outskirts of BEC, which is quite
different from the nonrotating case. In the presence of dissipation, the
steady visible vortex lattice in the BEC is mainly formed by the competition
between the rotating driving and the BB repulsive interaction. The BF
repulsion and the centrifugal force acting on the bosons tend to push the
fermionic superfluid toward the outside or the inside. Considering as well
the coherence of the fermionic superfluid with FF attractive interaction, it
evidently prefers to occupy the trap center, especially in the presence of a
central density hole for the bosons, because the corresponding potential
energy for the superfluid fermions is relatively small such that the system
energy of the SBFM reaches the minimum.

In Figs. 4(a1) and 4(a2), we show the density profiles of the BEC and the
fermionic superfluid in a static SBFM with $N_{b}=10^{3}$, $N_{p}=10^{5}$, $%
a_{b}=5$ nm, $R_{1}=6$, and $R_{2}=-20$. \ Compared with the layer-separated
phase in Figs. 2(a1) and 2(a2), here the torus density profile of the BEC
becomes thinner while the circular density distribution of the fermionic
superfluid gets larger. The steady structure at $t=500$ of the SBFM rotating
with $\Omega =0.92$ is displayed in Figs. 4(b1) and 4(b2), and the
corresponding phase profiles are given in Figs. 4(c1) and 4(c2). From Figs.
4(b1) and 4(c1), we can see that there are no visible vortices, but there
exist some hidden vortices in the BEC, where the trap center is occupied by
multiple close singly quantized hidden vortices. Here the central blurred
region in the phase profile mainly results from the inevitable numerical
errors or fluctuations in the numerical computations. In contrast to Fig. 3,
a triangular visible vortex lattice made of six visible vortices forms in
the rotating fermionic superfluid, as shown in Figs. 4(b2) and 4(c2).
Although there is attractive FF interaction between two fermions in a Cooper
pair, the fermionic superfluid can display a weak effective repulsive
interaction between the Cooper pairs under the appropriate parameters
according to Eqs. (\ref{SFE}), (\ref{ChemPotentialFermion}), (\ref%
{FermionFunction}), and (\ref{DimLess2DSFE}), which is consistent with the
relevant theoretical prediction \cite{Pieri}. This point may explain why
visible vortices can be generated in the rotating fermionic superfluid.

\begin{figure}[tbp]
\centerline{\includegraphics*[width=7.8cm]{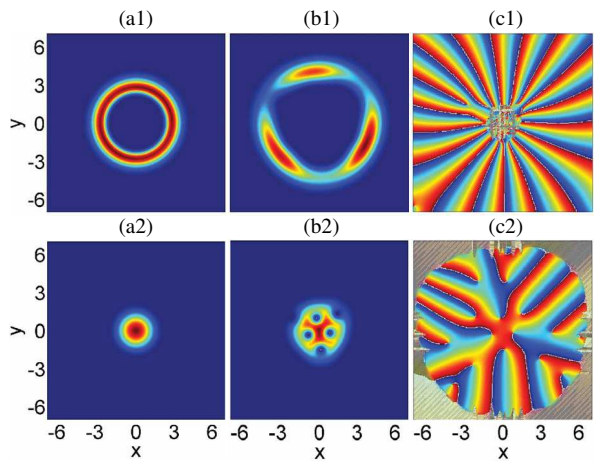}}
\caption{(Color online) (a1), (a2) Ground-state density profiles of a static
superfluid $^{7}$Li-$^{6}$Li mixture with $N_{b}=10^{3}$ and $N_{p}=10^{5}$.
(b1), (b2) The steady density profiles at $t=500$ for the superfluid
Bose-Fermi system rotating with $\Omega =0.92$. (c1), (c2) The corresponding
phase profiles. The relevant parameters are $a_{b}=5$ nm, $R_{1}=6$, and $%
R_{2}=-20$. Here, $1$ denotes the bosonic superfluid, $2$ represents the
fermionic superfluid, and $x$ and $y$ are in units of $d_{0}$. The darker
color area indicates lower density or phase. }
\label{Figure4}
\end{figure}

\section{Dynamics of vortex formation in a rotating SBFM}

In order to reveal the vortex formation process in the rotating superfluid
mixture of bosons and Cooper pairs, we consider a bigger superfluid
Bose-Fermi system which requires more computation effort. In Figs. 5(a1) and
5(a2), we present the ground-state structure of a static SBFM, where the
parameters are $N_{b}=10^{4}$, $N_{p}=10^{5}$, $a_{b}=50$ nm, $R_{1}=6$, and
$R_{2}=-0.1$. The steady structure at $t=500$ of the SBFM rotating with $%
\Omega =0.96$ is shown in Figs. 5(b1) and 5(b2), and the corresponding phase
profiles are displayed in Figs. 5(c1) and 5(c2). The initial state of the
SBFM at $t=0$ has a shell-shaped separated structure with the fermionic
superfluid being surrounded by the bosonic superfluid, which is similar to
that in Figs. 2(b1) and 2(b2). In the presence of dissipation, a steady
visible vortex lattice forms eventually in the outer of the superfluid BEC
[Fig. 5(b1)], where the energy of the rotating SBFM reaches the minimum in
the rotating frame. This point is similar to the case of Fig. 3(b1). The
large density hole of the bosons corresponds to 22 singly quantized hidden
vortices indicated by the phase profile in Fig. 5(c1). The central blurred
region in the phase profile is mainly caused by the inevitable numerical
errors or fluctuations in the numerical computations. In contrast to the
case of $N_{b}=1000\ $and $N_{p}=100$ (Figs. 2 and 3), seven evident visible
vortices are generated in the fermionic superfluid [see Figs. 5(b2) and
5(c2)].

\begin{figure}[tbp]
\centerline{\includegraphics*[width=8.4cm]{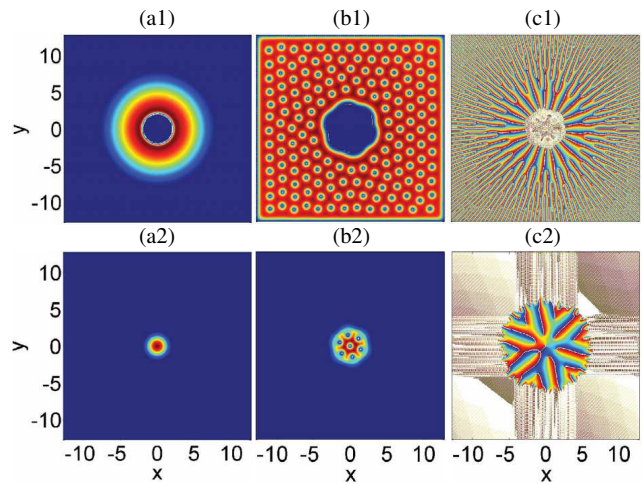}}
\caption{(Color online) (a1), (a2) Ground-state density profiles of a static
superfluid $^{7}$Li-$^{6}$Li mixture with $N_{b}=10^{4}$ and $N_{p}=10^{5}$.
(b1), (b2) The steady density profiles at $t=500$ for the superfluid
Bose-Fermi system rotating with $\Omega =0.96$. (c1), (c2) The corresponding
phase profiles. The relevant parameters are $a_{b}=50$ nm, $R_{1}=6$, and $%
R_{2}=-0.1$. Here, $1$ denotes the bosonic superfluid, $2$ represents the
fermionic superfluid, and $x$ and $y$ are in units of $d_{0}$. The darker
color area indicates lower density or phase. }
\label{Figure5}
\end{figure}

The dynamical evolution of the rotating SBFM is illustrated in Fig. 6, where
the top two rows denote the time evolutions of the density profiles of the
BEC (row 1) and fermionic superfluid (row 2), while the bottom two rows
represent those of the phase profiles of the BEC (row 3) and fermionic
superfluid (row 4). The evolution times are $t=11.5$ (left), $t=20$
(middle), and $t=60$ (right), respectively. Initially, the superfluid
densities $\left\vert \psi _{b}(x,y,t=0)\right\vert ^{2}$ and $\left\vert
\psi _{p}(x,y,t=0)\right\vert ^{2}$ in a stationary isotropic harmonic trap
are shown in Figs. 5(a1) and 5(a2). With the development of time, the
boundary surface of the bosonic superfluid undergo complex turbulent
oscillation and many ghost vortices appear at the outskirts of the BEC,
which can be seen in Figs. 6(a1) and 6(a3). In contrast, the fermionic
superfluid remains basically unchanged, as shown in Figs. 6(a2) and 6(a4).
Essentially, these ghost vortices are generated by collective excitations
through the nonlinear atomic interactions and the Landau instability
associated with the negative excitation frequency \cite{Kasamatsu1,Wu}
because the rotating harmonic trap is isotropic and has rotation symmetry.
This characteristic is evident especially for the component of fermionic
superfluid [see Figs. 6(a2)--6(c2) and Figs. 6(a4)--6(c4)]. Thus here the
formation mechanism of topological defects is different from the case of a
rotating anisotropic harmonic potential, where ghost vortices mainly result
from the dynamical instability through the rapid modulation of trapping
anisotropy \cite{Tsubota,Kasamatsu1,Dalfovo,Madison}. The angular momentum
is transferred into the SBFM via the excitations of surface modes or the
generation of visible vortices and hidden vortices. With the further time
evolution, some ghost vortices penetrate into the BEC and become visible
vortices or hidden vortices due to the Landau instability [see Figs.
6(b1)--6(c1) and Figs. 6(b3)--6(c3)], where the visible vortices arrange
themselves irregularly. When the rotating SBFM reaches an equilibrium state,
the visible vortices form a triangular lattice such that the energy of the
system approaches the minimum in the rotating frame. Compared with the
rotating BEC, the vortex (including ghost vortex and visible vortex)
formation in the rotating fermionic superfluid exhibits an evident
hysteresis effect, as shown in Figs. 6 and 5.

\begin{figure}[tbp]
\centerline{\includegraphics*[width=8.4cm]{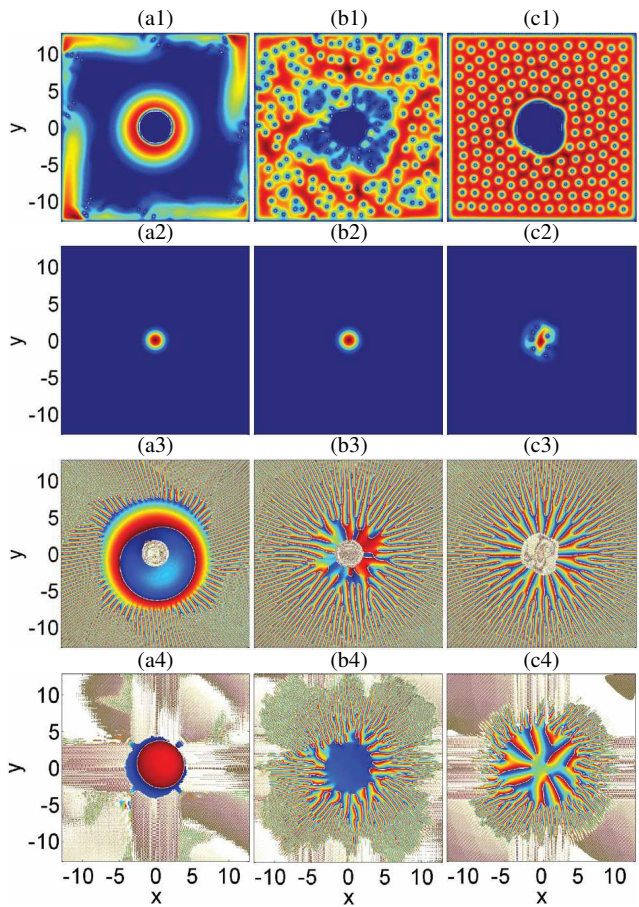}}
\caption{(Color online) Time evolutions of the density profiles $\left\vert
\protect\psi _{b}\right\vert ^{2}$ (row 1) and $\left\vert \protect\psi %
_{p}\right\vert ^{2}$ (row 2) and the corresponding phase profiles (rows 3
and 4) after the superfluid Bose-Fermi system rotates with $\Omega =0.96$,
where the other parameters are the same as those in Fig. 5. The value of the
phase varies continuously from $0$ to $2\protect\pi $. The time is (a1-a4) $%
t=11.5$, (b1--b4) $t=20$, and (c1--c4) $t=60$. The dark color area indicates
the lower density or phase. Here $x$ and $y$ are in units of $d_{0}$, and $t$
is in units of $1/\protect\omega _{bf}$.}
\label{Figure6}
\end{figure}

The structure and dynamics of a rotating SBFM are remarkably different from
those of rotating two-component BECs. First, in the latter case, the
vortices in the two BECs are always generated simultaneously \cite%
{Kasamatsu2,Woo,Mason}, while here the vortex formation in the superfluid
fermions is far later than that in the BEC due to the presence of Cooper
pair with attractive FF interaction. Second, in the presence of dissipation,
there are triangular visible vortex lattices and singly quantized hidden
vortices formed in a rotating SBFM. In our simulation, we did not observe
the generation of vortex sheet or square vortex lattice occurring in a
rotating two-component BEC \cite{Kasamatsu2,Mason,Kuopanportti,Ghazanfari}.
These structures as well as the giant vortex are suppressed by the
dissipation term in our simulation.

\section{Conclusion}

In summary, we have studied the structures for rotating and non-rotating
superfluid Bose-Fermi mixtures and the dynamics of vortex formation in a
rotating superfluid Bose-Fermi mixture. We show that the ratio of BF
interaction to BB interaction plays a key role in determining the
ground-state structure of a nonrotating SBFM. For fixed BB and BF $s$-wave
scattering lengths, the variation of FF (spin-up and -down) $s$-wave
scattering length does not influence the phase diagram for a nonrotating
SBFM. Furthermore, depending on the choice of parameters, we find that a
rotating SBFM with a sufficiently large angular velocity supports four
typical steady structures: a mixed phase and three layer-separated phases.
In addition, it is shown that different separated phases at the initial time
may result in almost the same steady structures. In particular, we find that
the generation of visible vortices in the fermionic superfluid exhibits an
evident hysteresis effect during the time evolution of a rotating SBFM. Our
results indicate that the topological structure and the dynamics of a
rotating SBFM are remarkably different from the cases of rotating
two-component BECs. We expect that our findings can be observed and tested
in future experiments. In the mean time, the present investigation provides
a way to further test the validity of the unitarity model \cite{Adhikari1}
for a SBFM.

\begin{acknowledgments}
We thank Biao Wu, Yongping Zhang, Li Mao and Yong Xu for helpful
discussions. L.W. acknowledges the research group of Professor Chuanwei
Zhang at The University of Texas at Dallas, where part of the computations
were carried out. This work was supported by the National Natural Science
Foundation of China (Grants No. 11475144, No. 11047033,and No. 11304270),
and the Ph.D. Foundation of Yanshan University (No. B846).
\end{acknowledgments}


\begin{thebibliography}{99}
\bibitem{Leggett} A. J. Leggett, \textit{Quantum Liquids} (Oxford University
Press, Oxford, 2006).

\bibitem{Pitaevskii} L. Pitaevskii and S. Stringari, \textit{Bose-Einstein
Condensation }(Oxford University Press, Oxford, 2003).

\bibitem{Pethick} C. J. Pethick and H. Smith, \textit{Bose-Einstein
Condensation in Dilute Gases}, 2nd ed. (Cambridge University Press,
Cambridge, 2008).

\bibitem{Zapf} V. Zapf, M. Jaime, and C. D. Batista, Rev. Mod. Phys. \textbf{%
86}, 563 (2014).

\bibitem{Liberati} S. Liberati and L. Maccione, Phys. Rev. Lett. \textbf{112}%
, 151301 (2014).

\bibitem{Rysti} J. Rysti, J. Tuoriniemi, and A. Salmela, Phys. Rev. B
\textbf{85}, 134529 (2012).

\bibitem{Ferrier-Barbut} I. Ferrier-Barbut, M. Delehaye, S. Laurent, A. T.
Grier, M. Pierce, B. S. Rem, F. Chevy, and C. Salomon, Science \textbf{345},
1035 (2014).

\bibitem{Ho1} T.-L. Ho and L. Yin, Phys. Rev. Lett. \textbf{84}, 2302 (2000).

\bibitem{YWu} Y. Wu, X. Yang, and C. Sun, Phys. Rev. A \textbf{62}, 063603
(2000).

\bibitem{Adhikari1} S. K. Adhikari and L. Salasnich, Phys. Rev. A \textbf{78}%
, 043616 (2008).

\bibitem{Adhikari2} S. K. Adhikari and L. Salasnich, Phys. Rev. A \textbf{76}%
, 023612 (2007).

\bibitem{Maeda} K. Maeda, G. Baym, and T. Hatsuda, Phys. Rev. Lett. \textbf{%
103}, 085301 (2009).

\bibitem{Ramachandhran} B. Ramachandhran, S. G. Bhongale, and H. Pu, Phys.
Rev. A \textbf{83}, 033607 (2011).

\bibitem{Abdullaev} F. Kh. Abdullaev, M. \"{O}gren, and M. P. Sorensen,
Phys. Rev. A \textbf{87}, 023616 (2013).

\bibitem{Wen1} L. H. Wen, H. W. Xiong, and B. Wu, Phys. Rev. A \textbf{82},
053627 (2010).

\bibitem{Recati} A. Recati, I. Carusotto, C. Lobo, and S. Stringari, Phys.
Rev. Lett. \textbf{97}, 190403 (2006).

\bibitem{Adhikari3} S. K. Adhikari, Phys. Rev. A \textbf{77}, 045602 (2008).

\bibitem{Lee1} T. D. Lee and C. N. Yang, Phys. Rev. \textbf{105}, 1119
(1957).

\bibitem{Lee2} T. D. Lee, K. Huang, and C. N. Yang, Phys. Rev. \textbf{106},
1135 (1957).

\bibitem{Blume1} D. Blume and C. H. Greene, \ Phys. Rev. A \textbf{63},
063601 (2001).

\bibitem{Carlson} J. Carlson, S.-Y. Chang, V. R. Pandharipande, and K. E.
Schmidt, Phys. Rev. Lett. \textbf{91}, 050401 (2003).

\bibitem{Astrakharchik} G. E. Astrakharchik, J. Boronat, J. Casulleras, and
S. Giorgini, Phys. Rev. Lett. \textbf{93}, 200404 (2004).

\bibitem{Blume2} D. Blume, J. von Stecher, and C. H. Greene, Phys. Rev.
Lett. \textbf{99}, 233201 (2007).

\bibitem{Chang} S. Y. Chang and G. F. Bertsch, Phys. Rev. A \textbf{76},
021603(R) (2007).

\bibitem{Salasnich} L. Salasnich, N. Manini, and F. Toigo, Phys. Rev. A
\textbf{77}, 043609 (2008).

\bibitem{Tsubota} M. Tsubota, K. Kasamatsu, and M. Ueda, Phys. Rev. A
\textbf{65}, 023603 (2002).

\bibitem{Kasamatsu1} K. Kasamatsu, M. Tsubota, and M. Ueda, Phys. Rev. A
\textbf{67}, 033610 (2003).

\bibitem{Wen2} L. H. Wen and X. B. Luo, Laser Phys. Lett. \textbf{9}, 618
(2012).

\bibitem{Zhang} Y. Zhang, L. Mao, C. Zhang, Phys. Rev. Lett. \textbf{108},
035302 (2012).

\bibitem{Xu} Y. Xu, Y. Zhang, B. Wu, Phys. Rev. A \textbf{87}, 013614 (2013).

\bibitem{Wen3} L. H. Wen, J. S. Wang, J. Feng, and H. Q. Hu, J. Phys. B
\textbf{41}, 135301 (2008).

\bibitem{Peaceman} D. W. Peaceman and H. H. Rachford, Jr., J. Soc. Ind.
Appl. Math. \textbf{3}, 28 (1955).

\bibitem{Wen4} L. H. Wen, Y. J. Qiao, Y. Xu, and L. Mao, Phys. Rev. A
\textbf{87}, 033604 (2013).

\bibitem{Choi} S. Choi, S. A. Morgan, and K. Burnett, Phys. Rev. A \textbf{57%
}, 4057 (1998).

\bibitem{Molmer} K. Molmer, Phys. Rev. Lett. \textbf{80}, 1804 (1998).

\bibitem{Roth} R. Roth, Phys. Rev. A \textbf{66}, 013614 (2002).

\bibitem{Capuzzi} P. Capuzzi, A. Minguzzi, and M. P. Tosi, Phys. Rev. A
\textbf{68}, 033605 (2003).

\bibitem{Adhikari4} S. K. Adhikari and L. Salasnich, Phys. Rev. A \textbf{75}%
, 053603 (2007).

\bibitem{Wen5} L. H. Wen, Y. P. Zhang, and J. Feng, J. Phys. B \textbf{43},
225302 (2010).

\bibitem{Ho2} T.-L. Ho and V. B. Shenoy, Phys. Rev. Lett. \textbf{77}, 3276
(1996).

\bibitem{Pu} H. Pu and N. P. Bigelow, Phys. Rev. Lett. \textbf{80}, 1130
(1998).

\bibitem{Trippenbach} M. Trippenbach, K. Goral, K. Rzazewski, B. Malomed,
and Y. B. Band, J. Phys. B \textbf{33}, 4017 (2000).

\bibitem{LWen} L. Wen, W. M. Liu, Y. Cai, J. M. Zhang, and J. Hu, Phys. Rev.
A \textbf{85}, 043602 (2012).

\bibitem{Adhikari5} S. K. Adhikari and L. Salasnich, New J. Phys. \textbf{11}%
, 023011 (2009).

\bibitem{Gubbels} K. B. Gubbels and H. T. C. Stoof, Phys. Rev. A \textbf{84}%
, 013610 (2011).

\bibitem{Mithun} T. Mithun, K. Porsezian, and B. Dey, Phys. Rev. A \textbf{89%
}, 053625 (2014).

\bibitem{Fetter} A. L. Fetter, Rev. Mod. Phys. \textbf{81,} 647 (2009).

\bibitem{Cooper} N. R. Cooper, Adv. Phys. \textbf{57}, 539 (2008).

\bibitem{Pieri} P. Pieri and G. C. Strinati, Phys. Rev. Lett. \textbf{91},
030401 (2003).

\bibitem{Wu} B. Wu and Q. Niu, Phys. Rev. A \textbf{64}, 061603(R) (2001).

\bibitem{Dalfovo} F. Dalfovo and S. Stringari, Phys. Rev. A \textbf{63},
011601(R) (2000).

\bibitem{Madison} K.W. Madison, F. Chevy, V. Bretin, and J. Dalibard, Phys.
Rev. Lett. \textbf{86}, 4443 (2001).

\bibitem{Kasamatsu2} K. Kasamatsu, M. Tsubota, and M. Ueda, Phys. Rev. Lett.
\textbf{91}, 150406 (2003).

\bibitem{Woo} S. J. Woo, S. Choi, L. O. Baksmaty, and N. P. Bigelow, Phys.
Rev. A \textbf{75}, 031604 (2007).

\bibitem{Mason} P. Mason and A. Aftalion, Phys. Rev. A \textbf{84}, 033611
(2011).

\bibitem{Kuopanportti} P. Kuopanportti, J. A. M. Huhtam\"{a}ki, and M. M\"{o}%
tt\"{o}nen, Phys. Rev. A \textbf{85}, 043613 (2012).

\bibitem{Ghazanfari} N. Ghazanfari, A. Kele\c{s}, and M. \"{O}. Oktel, Phys.
Rev. A \textbf{89}, 025601 (2014).
\end{thebibliography}
\end{document}